# Impact of Mixing on Flavor and Aroma Development in Fermented Foods

**Azarmidokht Gholamipour-Shirazi [a], Endre Joachim Lerheim Mossige [b]**

[a]*UK Future Biomanufacturing Research Hub, Manchester Institute of Biotechnology, School of Chemistry, University of Manchester*

[b]*RITMO Centre for Interdisciplinary Studies in Rhythm, Time and Motion, University of Oslo, Norway*

a.g.shirazi@manchester.ac.uk

joachim.mossige@imv.uio.no

## Abstract

*The flavor and aroma development in fermented foods is intricately tied to the mixing dynamics during fermentation. This review explores how variations in mixing influence the physical, chemical, and microbial interactions within fermentation systems, ultimately affecting sensory characteristics such as flavor and aroma. Factors like rheology, shear forces, and fluid flow patterns are critical in mass transfer, microbial activity, and the release of volatile compounds, contributing to fermented products' sensory profile. Examples from common fermented foods—including bread, yogurt, beer, wine, and cheese—highlight how controlled mixing can optimize the release of desirable flavor compounds, improve biosynthesis yields, and reduce technological complexity. Understanding these physical interactions is essential for advancing fermentation processes in the food industry, leading to higher product quality, better flavor retention, and enhanced consumer satisfaction.*

| Symbol | Definition | Units |
|---|---|---|
| $\mu$ | Dynamic viscosity of the solution | $Pa \cdot s$ |
| $\rho$ | Density of the fluid | $kg/m^3$ |
| $\tau$ | Shear stress | $Pa$ |
| $\dot{\gamma}$ | Shear rate | $s^{-1}$ |
| $\delta_D$ | Width of the concentration layer | $m$ |
| $D$ | Diffusivity of taste molecules | $m^2/s$ |
| $j$ | Molecular flux per unit area of the tongue | $mol/s$ |
| $v$ | Flow velocity | $m/s$ |
| $t_m$ | Mixing time | $s$ |
| $K$ | Consistency index (for power-law fluids) | $Pa \cdot s^n$ |
| $n$ | Flow behavior index (for power-law fluids) | *Dimensionless* |
| $Pe$ | Péclet number | *Dimensionless* |
| $Re$ | Reynolds number | *Dimensionless* |
| $L$ | Characteristic length (e.g., width of impeller) | $m$ |
| $\Delta c/\Delta x$ | Concentration gradient | $mol/m^3 \cdot m$ |
| $c_0$ | Concentration in the bulk | $mol/m^3$ |
| $k_L$ | Liquid-side mass transfer coefficient | $m/s$ |
| $A$ | Interfacial area for mass transfer | $m^2$ |
| $V$ | Volume of the bioreactor | $m^3$ |
| $Q$ | Flow rate | $m^3/s$ |

## Introduction

Fermentation, a cornerstone of food production, is instrumental in creating diverse flavors and textures and preserving food products. This ancient technique involves the metabolic breakdown of organic compounds by microorganisms, typically yeasts or bacteria, under anaerobic conditions.[1] In food production, fermentation serves multiple purposes: it enhances nutritional value, improves digestibility, extends shelf life, and develops unique sensory characteristics. Common fermented foods include bread, cheese, yogurt, wine, beer, and pickled vegetables. The process not only creates desirable flavors and aromas but also often increases the bioavailability of nutrients and produces beneficial compounds such as probiotics.[2]

In recent years, there has been renewed interest in fermentation due to its potential health benefits and its role in creating novel, artisanal food products that excite and intrigue the palate.[3] At its core, fermentation is a complex interplay of biochemical reactions, microbial growth, and physical processes that collectively shape the final product's sensory and nutritional characteristics.[4,5] Modern industrial fermentation processes in the food and beverage industry are remarkably complex and can be described from various perspectives. These include the type of bioreactor used (batch, fed-batch, or continuous mode of operation), the immobilization of the biocatalyst (free or immobilized cells/enzymes), the state of matter in the system (submerged or solid substrate fermentations), the type of culture used (single strain or mixed culture processes), the method of mixing in the bioreactor (mechanical, pneumatic, or hydraulic agitation), and the availability of oxygen (aerobic, microaerobic, or anaerobic processes). Each of these perspectives offers a unique insight into the fermentation process.[6] Among the various physical factors influencing fermentation, mixing plays a crucial role in determining the course and outcome of the process.

Mixing is a pivotal element in fermentation processes, influencing the distribution of substrates, microorganisms, and metabolites throughout the medium. This distribution directly impacts reaction rates, mass transfer, and overall fermentation kinetics.[7] Ni et al.[8] investigated mass transfer in a pulsed baffled bioreactor and a stirred tank fermenter using yeast cultures of varying concentrations and ages. In the pulsed baffled reactor, they observed complex eddy mixing with vortices forming behind baffles and moving toward the stream center during flow reversal. This efficient mixing eliminated dead zones and increased oxygen mass transfer by 75% compared to the stirred tank fermenter. The improvement was linked to enhanced interfacial area, elevated gas hold-up, uniform shear distribution, and thinner liquid films, leading to a higher liquid-side mass transfer coefficient.[8]

Mixing is not just about homogenization but the intricate dance of altering fluid dynamics within the mixture. It introduces variations in shear forces, flow patterns, and turbulence, adding a layer of intricacy to the fermentation process.[9,10] The interaction between fluid dynamics and mixing processes profoundly impacts the effective release of volatile compounds and the integration of flavors. These changes in fluid mechanics can significantly affect the development and release of flavor and aroma compounds, making it crucial to understand these physical interactions to optimize fermentation processes.

This mini-review aims to explore these aspects, focusing on how the physics of mixing influences the sensory characteristics of fermented foods (Fig.1). This review examines the intricate relationship between mixing dynamics and the development of flavors and aromas in fermented foods. We suggest that the rheological changes induced by different mixing regimes profoundly influence the sensory profile of fermented products by modulating microbial activity, substrate availability, and the formation and retention of volatile compounds. By examining the physics of fluid flow in fermentation systems, we aim to elucidate how mixing parameters such as shear rate and vorticity affect the complex biochemical processes underlying flavor and aroma development. This understanding is crucial for optimizing fermentation processes and tailoring the sensory attributes of fermented foods to meet consumer preferences and quality standards.

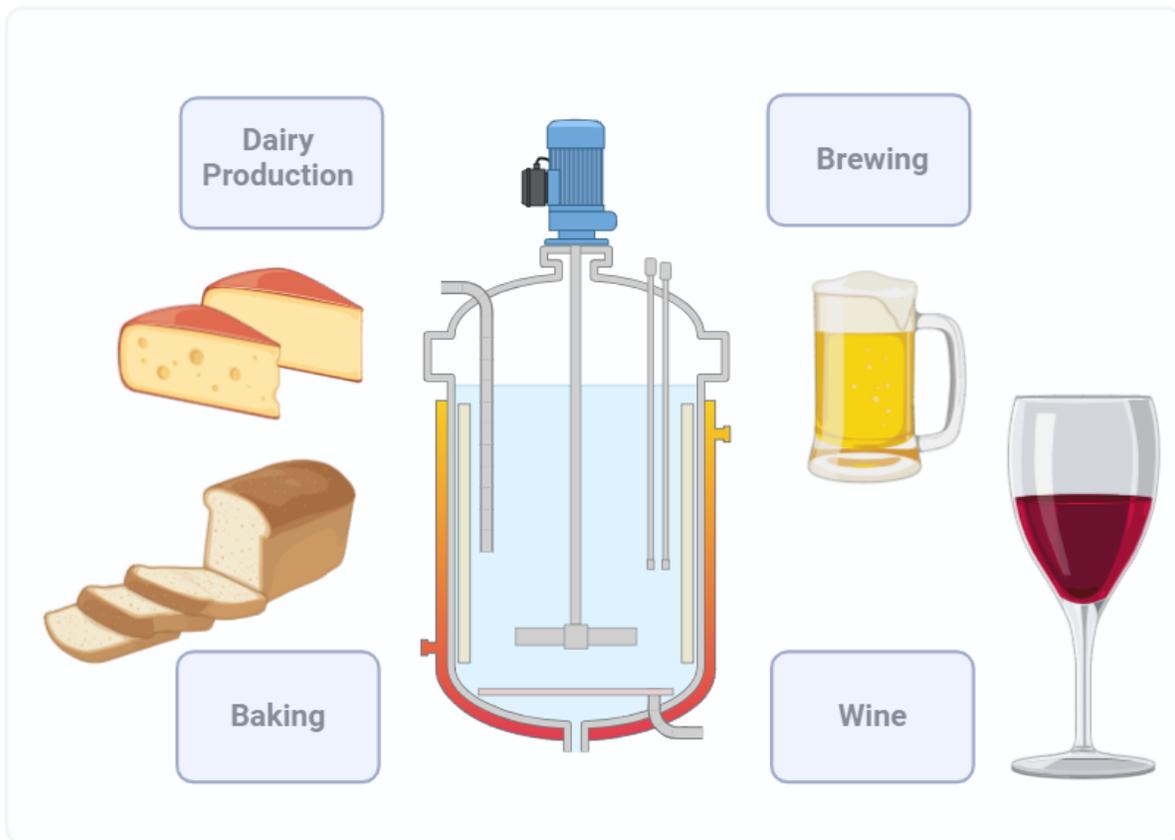

*Figure 1. The Interplay of Mixing, Fluid Flow, and Fermentation in Beer, Bread, Cheese, and Wine Production.*

## Rheology in Flavor and Aroma Development

The investigation into the relationship between rheology [11,12] and perception is known as psychorheology [13,14] and has a rich and extensive history. [15] Two primary objectives drive this field of study[16]:

(i) Predicting perception from instrumental rheological measurements: Utilizing instruments to predict sensory perceptions offers significant practical advantages over relying solely on sensory panels. This approach allows for more consistent, reproducible, and efficient evaluations of material properties.

(ii) Understanding the relationship between material structure and perception: Rheology is crucial in elucidating how physical and structural properties influence sensory experiences.[17] It is a vital tool in linking materials' mechanical and viscoelastic properties to their perceived texture and other sensory attributes, thereby enhancing our understanding of the sensory characteristics of fermented foods.

Rheological property such as viscosity, yield stress, shear thinning, and thickening is closely related to the sensory properties of food, such as mouth feel [18] and the intensity of flavor.[19] In particular, viscosity influences other taste perceptions, including saltiness, sweetness, bitterness, flavor, and "sting".[20] In highly viscous and firm gel systems, the availability of sodium ions appears to limit salt taste perception. However, texture-related interactions between taste and sensory cues significantly reduce the perceived saltiness at lower polysaccharide concentrations, especially when solid particles are present.[21] The increase in viscosity impacts the free water available in the solution, resulting in a decrease in sweetness intensity and, therefore, a reduction in flavor intensity.[22] Studies have shown that hydrocolloid solutions with more pronounced shear-thinning behavior tend to have a smaller impact on reducing sweetness perception compared to those with weaker pseudoplastic properties.[23] The presence of yield stress reflects greater structural integrity within a fluid, potentially aiding in forming a cohesive bolus and facilitating more efficient swallowing. Yield stress has been suggested as a key factor in a fluid's ability to be swallowed, acting as a force that must be overcome to propel the bolus from the mouth to the pharynx. However, in a study[24], a trained sensory panel detected significant differences in perceived propulsion effort across twelve thickened fluid samples. Still, these perceptions did not correlate with the presence or magnitude of yield stress. This suggests that yield stress is a poor indicator of oral cohesiveness and swallowing effort.

A long history of studies shows that if a solution's viscosity is increased, there is a reduction in perceived flavor or taste.[25] The viscosity can be increased by adding a hydrocolloid, for example. Ferry et al.[26] studied the effect of hydrocolloid type on flavor perception in aqueous systems and reported a considerable decrease in the perception of flavor and saltiness with increasing viscosity for xanthan and hydroxypropylmethylcellulose (HPMC)- thickened products.

The reason why enhanced viscosity decreases the perception of taste is that it slows down the transport of taste molecules to the receptors on the tongue.[25] For a molecule to reach a taste receptor, it is first advected by the flow inside the mouth, and then it must diffuse down a thin concentration boundary layer covering the tongue, see Fig. 2. The width of this concentration layer (which is typically about ten times thinner than a flow boundary layer) depends on the downstream distance along the tongue $x$ and the relative importance of convective to diffusive transport of molecules, given by the Péclet number, $Pe = \dot{\gamma} x^2 / D$, where $\dot{\gamma}$ is the shear rate of the flow, and $D$ is the diffusivity of taste molecules[27]:

$$\delta_D \sim x Pe^{-\frac{1}{3}} \sim x \left(\frac{D}{\dot{\gamma} x^2}\right)^{\frac{1}{3}} \sim D^{\frac{1}{3}}. \qquad \text{Eq. 1}$$

Note that the concentration layer follows a different scaling relation than the momentum boundary layer, whose thickness scales as $\delta \sim x Re^{-1/2} \sim \sqrt{\nu/\dot{\gamma}}$, where $Re = x^2 \dot{\gamma}/\nu$ is the Reynolds number and $\nu$ is the kinematic viscosity.

If the taste molecules are far apart so that they do not interact, the diffusivity of a single taste molecule inside the concentration boundary layer is inversely related to the solution viscosity, $\mu$, via the Stokes-Einstein relation[28]:

$$D \sim \frac{1}{\mu}. \qquad \text{Eq. 2}$$

As such, the molecular flux per unit area of the tongue, which controls the perceived taste, is

$$j = D\frac{\Delta c}{\Delta y} \sim D\frac{c_0}{\delta_D}, \qquad \text{Eq. 3}$$

where $\Delta c/\Delta y$ is the concentration gradient, and $c_0$ is the concentration in the bulk. By plugging Eq. 1 and Eq. 2 into Eq. 3, one obtains the dependency of flux on the solution viscosity:

$$j \sim \mu^{-2/3} c_0. \qquad \text{Eq. 4}$$

Put in words, the number of taste molecules reaching the receptors per unit time drops down by a factor of -2/3 with viscosity, leading to less perceived flavor. Intriguingly, to the best of our knowledge, the exact relation between flux and perceived taste is not known and could be an exciting new avenue of food science research.

Finally, one can estimate the influence of viscosity on the time it takes for a taste molecule to cross a concentration layer of thickness $\delta$:

$$\tau_D \sim \frac{\delta^2}{D} \sim \frac{\left(\mu^{-1/3}\right)^2}{\mu^{-1}} \sim \mu^{1/3} \qquad \text{Eq. 5}$$

In fermentation broths, the solution is often thickened by polymers that are not well mixed with the solution containing taste molecules. The perceived flavor may in this case be additionally reduced as long polymer molecules entangle and create a highly viscous layer covering the receptors.

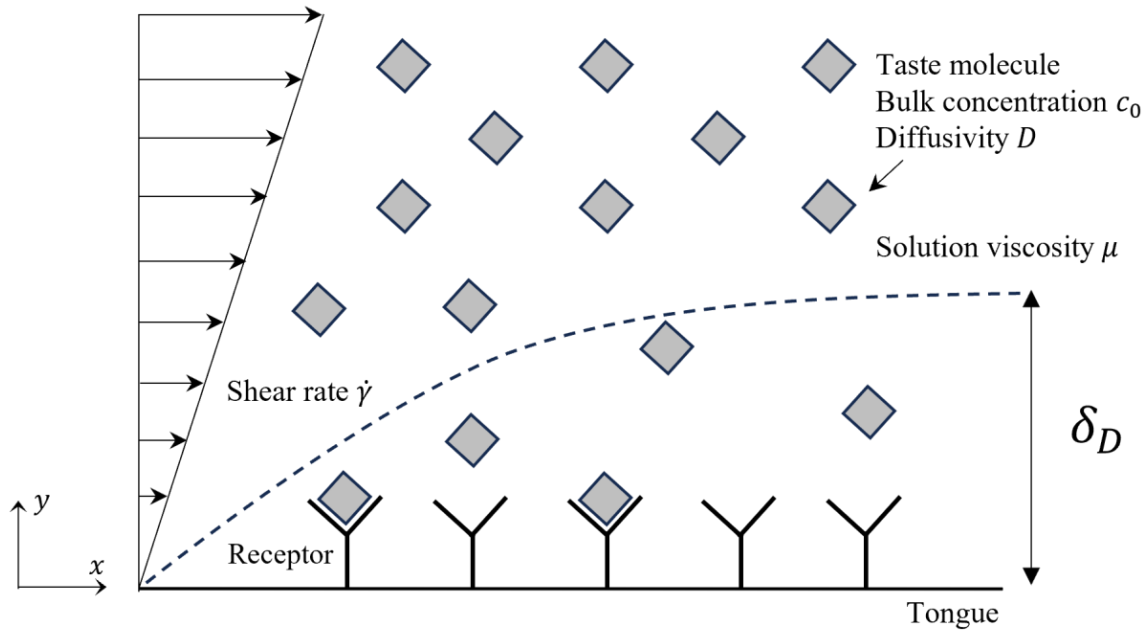

*Figure 2. For a taste molecule to reach a receptor, it must first be advected by the flow and then diffuse down a concentration layer with thickness $\delta_D$. The viscosity of the solution slows down the flux of molecules to the receptors, thereby reducing the perceived taste.*

## Mixing and Fermentation

The rheological behavior of the fermentation broths is one factor that decisively influences the performance of the biosynthesis processes. The medium viscosity combined with a non-Newtonian fluid affects the mass and heat transfer, mixing intensity, cell growth rate, and product accumulation rate.[29] Besides the high viscosities of fermentation broths, values reached in the chemical industry only for polymeric melt or solutions, the fermentation processes are characterized by the modification of the broths' rheology in the course of the biosynthesis cycle as the result of the biomass or high molecular weight products (extracellular polysaccharides, proteins, enzymes) accumulation and/or substrate consumption. For this reason, the values of the rheological parameters for fermentation liquids (viscosity, shear stress, shear rate, power law index, consistency index, yield stress) indicate the biosynthesis stage, a modality for controlling the process development.[30] Karsheva et al.[31] discuss previous studies on the relationship between fermentation time and rheological parameters. These studies highlight sinusoidal variations in yield stress and plastic viscosity over time, along with a peak in the consistency index $K$ and a dip in the flow index $n$. These patterns are linked to morphological changes in the fermentation broth, where microorganisms transition between pellet and filamentous forms. During fermentation, the average shear stress at a constant shear rate increases over time. This corresponds to a growth-rate curve with two peaks, reflecting the transition of mycelia between these structures. The trough between the peaks indicates the breakdown of the pellet form. This structural shift offers one possible explanation for the sinusoidal behavior of the rheological parameters. Another explanation is the interplay between biomass growth and apparent viscosity, where viscosity decreases as the lysis process begins.

Each microbial culture medium has a specific predominant rheological behavior, combined or not with other flow types, as a function of biosynthesis conditions. The rheological parameters influence decisively the performances of the fermentation, controlling the rate of mass and heat transfer, mixing, aeration, and downstream processes (filtration, extraction).

Sometimes, modifying the main rheological behavior is possible with the development of the biosynthesis process or the function of the operating conditions. Thus, it was observed that the fermentation broths of *S. griseus* initially exhibit a pseudoplastic flow. However, the liquid becomes Bingham plastic type for higher levels of mixing intensity, respectively, and for shear rate values over 5 s$^{-1}$.[32] For fungus fermentation broths, the morphology and concentration of microorganisms decisively influence the medium rheology. In function of the cultivated strain and fermentation conditions (airflow, mixing intensity, composition, and concentration of nutritive media), the fungus can grow in two morphological forms: filamentous biomass or agglomerated biomass (pellets). The morphological structure controls the rheological characteristics of the broths.[33] Still, irrespective of the biomass morphology type, the microorganisms' accumulation modifies the value of the rheological parameters or the rheological behavior.

The mixing efficiency in bioreactors is quantitatively described using the mixing time, which is the time it takes to obtain a certain homogeneity of an added tracer such as a food dye, starting from a completely segregated situation. Traditionally, the mixing time has been defined as

$$t_{mix} = \frac{c}{u}, \qquad \text{Eq. 6}$$

where $u$ is the velocity of the impeller, see Fig. 3, and $c$ is an experimentally defined constant. Bisgaard et al.[34] suggest using the so-called "retention time" instead, which is the time it takes for a fluid element to circulate in the bioreactor:

$$t_{retention} = \frac{V}{Q}, \qquad \text{Eq. 7}$$

where $V$ is the volume of the bioreactor, and $Q$ is the flow rate. While neither Eq. 6 nor Eq. 7 incorporates molecular diffusion, which is necessary for mixing, mixing is in general due to both convection (flow) and diffusion, and their relative importance is given by the Péclet number:

$$Pe = \frac{conventive\ transport}{diffusive\ transport} = \frac{\dot{\gamma} s_0^2}{D}. \qquad \text{Eq. 8}$$

Here, $D$ is the diffusivity of the tracer with initial blob size of $s_0$, and $\dot{\gamma}$ is the shear rate. The mixing time depends on the shear rate and the Péclet number, as well as the flow field. For example, in a two-dimensional linear shear flow, the mixing time scales as[35]

$$t_{mix} \sim \frac{1}{\dot{\gamma}} Pe^{1/3} \qquad \text{Eq. 9}$$

while in a chaotic flow with exponential stretching of fluid elements such as a tracer, which is more representative of mixing processes in bioreactors, the mixing time scales as[35]:

$$t_{mix} \sim \frac{1}{\dot{\gamma}} \log(Pe) \qquad \text{Eq. 10}$$

In regions far from the stirrer, the shear rate dramatically diminishes, which leads to an increase in the local mixing time.

An increase of broth viscosity and accentuation of non-Newtonian behavior also leads to a significant increase in the mixing time. This is because filament stretching is less efficient in such fluids[30], and since to the diffusivity is inversely related to viscosity, as seen in Eq.1. Fortunately, using efficient mixing, the negative effects of the increase in broth viscosity and the modifications of the rheological behavior during the fermentation cycle can be reduced without neglecting the impact of the shear rate on cultivated microorganisms. The combined effects of non-Newtonian behavior and inefficient mixing generally lead to reduced biosynthesis yields, challenges in analyzing the fermentation system, ineffective process control, increased technological costs, and a more complex system design.

When mixing non-Newtonian broths with high apparent viscosity, a zone with elevated shear stresses—known as the so-called "cave region"—forms around the stirrer, distinct from the bulk of the medium, see Figure 3 (left)[30]. The extent of the cave region is difficult to predict due to the complex rheology of fermentation broths. However, in cases where the rheology is relatively independent of shear rate, and the velocities parallel to the stirrer are much larger than the velocities perpendicular to it, the extent of the cave region can be approximated as the width of the flow boundary layer, $\delta_{stirrer}$ [36]

$$\delta_{stirrer} \sim \frac{L}{\sqrt{Re}}, \qquad \text{Eq. 11}$$

where $L$ is the width of the impeller, and $Re$ is the Reynolds number:

$$Re = \frac{uL}{\nu} = \frac{NL^2}{\nu}. \qquad \text{Eq. 12}$$

Here, $u$ is the impeller velocity, $N$ is the impeller frequency and $\nu$ is the kinematic viscosity.

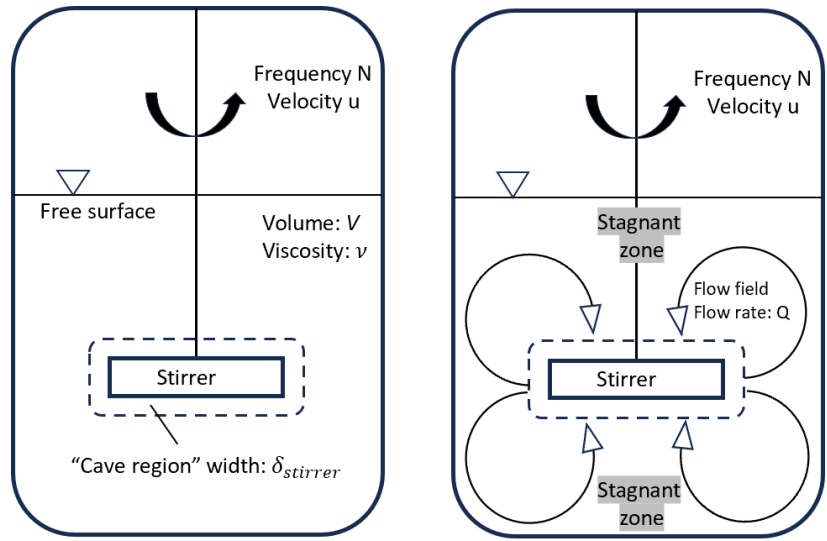

*Figure 3. Working principle of a stirred bioreactor. Left: A cave region characterized by high shear stresses forms around the stirrer. The mixing efficiency reaches a maximum in this region. Right: Stagnant zones with poor mixing efficiency develop far from the stirrer.*

In the cave region, the shear rates reach a maximum, and this leads to maximal mixing efficiencies. However, outside the cave region, the mixing efficiency decreases rapidly since the local shear rate diminishes rapidly with distance from the impeller. The mixing efficiency is particularly impaired in stagnant zones, which develop due to the no-slip condition at the container walls and at the fluid-gas interface (free surface), see Fig.3 (right). In stagnant zones, the substrate concentration is lower than the value needed for microorganism metabolism. For example, for *S. niveus* cultures, the apparent viscosity decreases with an increase in shear rate, i.e., shear thinning fluid, thus increasing with the distance from the stirrer. Therefore, air bubbles will rise in low viscosity regions near the stirrer, and as such, the medium will be more oxygenated there.

The role of the stirrer is not just to mix fluids; an equally important role is to aerate the liquid by forming air bubbles. The mass transfer of oxygen scales with the total area of the bubbles; as such, it is maximized by minimizing the size of individual bubbles, and by avoiding bubble coalescence. In non-Newtonian liquids, increased drag forces accelerate bubble coalescence, leading to the formation of larger bubbles. This coalescence reduces the effective interfacial area, which in turn decreases gas holdup in these viscous systems.[37] Finally, it is worth noting that for the same specific power consumption for broth mixing, the mass transfer coefficients of oxygen or other substrates are lower in non-Newtonian broths than in Newtonian broths.[38]

In the following sections, we will examine the impact of mixing during fermentation on the flavor and aroma of bread, wine, vinegar, beer, and yogurt.

## Bread

The bread-making process can be broadly classified into three sub-processes: mixing, fermentation, and baking. The mixing and fermentation processes play an essential role in determining the qualities of white bread. The mixing process also affects the bread's texture, influencing the perception of flavors. A well-mixed dough leads to a light and airy crumb, which can enhance the perception of flavors by releasing more flavor compounds upon chewing.

The mixing step, commonly known as "kneading," is essential for blending and homogenizing the ingredients (typically wheat flour, water, yeast, salt, and other minor components) in precise proportions while incorporating air bubbles to create a uniform dough. The gas bubble structure created during the mixing stage expands further with the release of carbon dioxide during progressive fermentation, making the dough less resistant to deformation, which is needed later to form a consistent cell structure in the finished bread.[39] This is predominantly made possible by the physical development of gluten through the hydration and interactions between the mixing components.[40] During mixing, the gluten proteins can be linked via disulfide bonds, hydrogen bonding, and hydrophobic interaction to form cross-links within peptide chains.[41] [42,43] When mixed with water, gluten proteins (gliadins and glutenins) uniquely form a robust and viscoelastic network, unlike any other plant protein. This network is primarily created through disulfide linkages between the amino acid cysteine and other noncovalent bonds. This structure plays a crucial role in determining the baking quality of bread by imparting the dough with water absorption capacity, cohesiveness, viscosity, and elasticity.[43] Both gliadin and glutenin fractions are required to form a viscoelastic network in the dough.[44] Gliadins contribute viscosity and extensibility to the dough, enabling it to rise during fermentation.[45] while glutenins impart tenacity and strength, preventing the dough from overextending and potentially collapsing during fermentation or baking[46–48]. On a microscopic scale, during the initial stage of protein hydration, fibrils of hydrated protein begin to interact and attach, forming a coarse primary network of large strands. Mixing stretches and thins these strands, allowing them to interact and align according to the stretching direction. At optimal mixing conditions, the protein fibrils interact two-dimensionally rather than just along the axes of individual strands, forming a continuous gluten network (gluten sheet). The extent of this network indicates that the dough has been sufficiently mixed.[49]

The rheological properties of dough are good indicators of its mixing, sheeting, and baking performances. Dough expansion and bread texture depend on dough rheological characteristics, primarily due to the gluten and starch components.[50] The complex nature of the dough and the final bread product, which is a multiphasic system, presents a significant challenge in understanding its behavior during breadmaking. Despite the participation of dough rheology in every stage during breadmaking, rheological principles are not applied to the production procedure. The rheological behavior of dough, and later as a bread crumb, is generally considered nonlinear. This nonlinearity necessitates the separate assessment of the viscous and elastic components with strain and strain rate and the need for consistent and homogeneous sample shapes and loading patterns to characterize the dough's rheological properties.[51,52]

Dough undergoes several types of deformation during each step of breadmaking. At the mixing stage, extreme deformation is exerted on the dough beyond the rupture limits. Much smaller deformations occur during fermentation, and deformation is intermediate during the subsequent step of sheeting and shaping. More deformation takes place at the proofing and baking stages.[47] However, most of the rheological properties of dough are decided during the early steps of production, particularly kneading and fermentation, as a result of the interactions that occur among the primary ingredients (flour, water, yeast, and air). Among these, high extensibility during baking and viscosity are argued to be the two critical rheological properties sufficient in translating into an excellent final bread product with high viscosity for preventing gas cells from rising and high extensibility for preventing sudden breakage in gas cell membranes.[52] Extensibility is enhanced significantly with a structure high in glutenin, while all dough essentially possesses adequate viscosity.[53–55] Dough rheology changes dramatically during baking as the elastic components of the dough start to dominate. This is believed to be controlled by the change in gluten rheology, especially within the temperature range of 55 to 75 °C.[56] is related directly to the amount of starch in the system available for gelatinization.[57]

## Wine and Vinegar

Wine aroma is a crucial indicator of wine quality, significantly influencing its hedonic appeal. The chemical composition of wine is the foundation of the sensory response and is determined by many factors. These include the grape variety, the geographical and viticultural conditions of grape cultivation, the microbial ecology of the grape and fermentation processes, and winemaking practices.[58] This complexity varies even with aging and bottling.[59] The study of wine aroma compounds is challenging due to their ultra-low concentrations, yet it offers opportunities for quality refinement.[60] Importantly, research has shown that the overall flavor compounds detected analytically do not directly correlate with perceived sensations during consumption. Instead, non-volatile components of the wine matrix play a crucial role in aroma perception and release, interacting with specific volatiles to influence the wine's sensory characteristics.[61]

Despite the importance of wine tasting, using a sensory panel can be expensive, and the training can take longer than instrumental characterization. Also, it is possible that others may not understand the terms used by an expert with special sensory training. Furthermore, as panelists are trained or specialized in particular products or sets of products, perceptions of the wine body can vary significantly. One expert might perceive a "heavy" wine as "light," while another expert trained in a different wine-producing region. This discrepancy in perception makes cross-regional comparisons challenging, as the frame of reference for wine characteristics can differ based on the panelist's background and expertise.[62] Therefore, if wine mouthfeel could be quantitatively measured using an instrumental technique, that may allow wineries to have a faster, repeatable, harmonized, and cheaper characterization complementary to the use of a panel of experts. The rheological behavior of wines has yet to be extensively discussed and only a few works have evaluated its basic rheological properties. Kosmerl et al.[63] studied the density and viscosity dependence on temperature for 40 Slovenian wines' for temperatures ranging between 20 and 50 °C. They found the density to be weakly non-linear function of temperature, $T$, well described by:

$$\rho(T) = \rho(T_0) + a_1 T - T_0 + a_2 T - T_0^2 \qquad \text{Eq. 13}$$

where $a_1$ and $a_2$ are experimental fitting parameters, and $T_0$ = 25 °C. The authors found the viscosity to decay exponentially with temperature, following the functional forms below:

$$\eta(T) = \exp(A - BT - CT^2) \qquad \text{Eq. 14}$$

$$\eta(T) = \exp(D - ET - FT) \qquad \text{Eq. 15}$$

where $B, C, D, E, F$ are fitting constants.

In another study[64], wines from the grapes collected in the Czech Republic were examined for their rheological attributes for temperatures between 5 and 40 °C. All eight wines were shear-thinning at low temperature (5 °C), and seven of them showed thixotropic behavior at this temperature as well. However, at temperatures above 10 °C, all eight wines behaved as Newtonian fluids. The authors speculate that the shear-thinning behavior at low temperature is due to attractive forces between suspended particles. The relative importance of such forces decreases with increased temperature since Brownian (random) fluctuations increase with temperature.

Density and viscosity are properties that significantly influence the body of wines. Neto et al. [65] studied the effect of ethanol, dry extract, and reducing sugars on the density and viscosity of Brazilian red wines. In accordance with the general behavior of liquids, they found that the wine viscosity and density decreased with increasing temperature, and as expected for ethanol-water mixtures, the density decreased with wine alcohol content as ethanol is lighter than water. Furthermore, they found that all the nine wines in their study behaved as Newtonian liquids for temperatures between 2 and 26 °C and shear rates between 1 and 250 $s^{-1}$, and they found the Arrhenius model to describe the viscosity dependence on temperature well:

$$\mu(T) = \beta \exp\left(\frac{G}{RT}\right) \qquad \text{Eq. 16}$$

Here, $\beta$ is an experimental fitting constant, $G$ is the universal gas constant ($8.314\ J\ mol\ K^{-1}$), $T$ is absolute temperature ($K$), and $G$ is the activation energy ($J\ mol^{-1}$). Finally, they found the viscosity to increase with the concentration of suspended, non-soluble particles, known as the dry extract. This agrees with Einstein's widely celebrated relation[66] between viscosity and particle volume ($\phi$):

$$\mu(\phi) = \mu\left(1 + \frac{5}{2}\phi\right) \qquad \text{Eq. 17}$$

Note that while the equation above was developed for hard spheres and in the dilute limit ($\phi \ll 1$), where both short-range interactions (e.g. steric, electrostatic) and long range (hydrodynamical) interactions are negligible, formulations more applicable to real fluids with higher volume fractions and different particle shapes exist, see e.g. Guazelli & Morris[67]. For a thorough review on the rheology of food suspensions, see Genovese et al.[68]. Finally, if soluble ingredients are added to wine, the viscosity may actually decrease, which is the opposite response one should expect if one adds insoluble ingredients. For example, Wang et al.[69] added a cocktail containing ingredients such as sodium sulphite, tartaric acid and tannins into red wine and found the viscosity to increase after storing the wine for 24h.

While rheology (and tribology) techniques offer valuable insights into the physical properties of wines, they present challenges in directly correlating with sensory perception. Although rheological measurements can effectively differentiate between wine samples, the subtle changes detected by these instruments often exceed the sensitivity of human oral perception. The study by Laguna et al.[62] demonstrates that while instrumental techniques can capture fine distinctions in wine texture, these differences may not always translate to noticeable variations in mouthfeel as consumers experience them.

Finally, it is worth noting that vinegar can be used as a model system for wine as its rheological properties can serve as a simple and reliable quality indicator. Various factors impact these properties, such as sugar content, sodium chloride, polyphenols, and the polymerization that occurs during aging. Glucose concentration and the glucose-to-fructose ratio significantly affect vinegar's steady shear viscosity and flow activation energy.[70,71] The flow behavior index is a measure of the degree of non-Newtonian behavior, and for vinegar, it is inversely related to the concentrations of glucose, fructose, and the °Bx (degrees Brix) level. Polymerization during extended aging alters vinegar's rheological properties, shifting its behavior from purely viscous to shear-thinning.[70,72,73]

Beer

Traditionally, brewing fermentations have been carried out without mechanical agitation, relying solely on the natural fluid motion generated by the $CO_2$ released during the batch process. This practice has persisted in the industry, mainly due to the concern that rotating agitators might harm the yeast.[74]

Brewers, especially those producing high-volume lager beers, focus on minimizing costs by optimizing raw materials and reducing process time. One approach to shorten primary fermentation is to increase fermentation temperature, which can reduce fermentation time but negatively impact flavor balance due to changes in yeast growth and biochemical pathways.[75] Another method is increasing the pitching rate (yeast concentration), which can reduce fermentation time without significantly affecting flavor.[75] Additionally, carbon dioxide pressure affects fermentation. Lower pressures stimulate yeast growth, while higher pressures inhibit growth without affecting sugar consumption.[76]

The most well-described flavor-active esters in beer are ethyl acetate (solvent-buttery-like aroma), ethyl caproate, ethyl caprylate (sour apple-like flavor and aroma), isoamyl acetate (fruity, banana aroma), isobutyl acetate, phenyl ethyl acetate, and ethyl octanoate (honey, fruity, roses, flowery aroma).[77] Research[78] has demonstrated a positive correlation between the specific power input (power input per unit volume of liquid) in a brewing vessel, which controls the degree of mixing, and the concentration of flavor compounds. Given that mixing is influenced by the rate of $CO_2$ evolution, liquid depth, vessel size, and aspect ratio, these factors can significantly impact the beer's flavor profile. This interaction has important implications for producing a consistent product across different locations where varying vessels are used.[78]

Boswell et al. [79] found that impeller speed has opposite effects on the formation rate, maximum concentration of esters, and higher alcohols. They observed that increased specific power input and turbulence from higher impeller speeds positively impacted the formation of higher alcohols, such as isobutanol and isoamyl alcohol. However, the formation of esters like ethyl acetate, ethyl hexanoate, and isoamyl acetate was suppressed at higher impeller speeds. Interestingly, for isobutyl acetate, the negative impact of agitation was not evident; in fact, agitation appeared to enhance its formation. This suggests the metabolic pathways leading to isobutyl acetate formation might be uniquely enhanced under agitated conditions.

Yogurt and Cheese

Yogurt is commercially produced through fermentation by lactic acid bacteria (commonly Lactobacillus spp. and Streptococcus spp.) at temperatures usually 27 to 40°C. The most traditional method for yogurt preparation involves milk fermentation by lactic acid bacteria in the same containers where the product will be distributed. In traditional yogurt manufacturing, the yogurt is not agitated during fermentation. The use of non-stirred small vessels is also possible. Mixing is applied after fermentation for liquid formulations or formulations mixed with fruit. However, stirring could improve heat and mass transport across the fermentation tank. Three primary arguments have been made against using stirring during the fermentation process in industrial yogurt production: (a) stirring could disrupt the gel formation, which is crucial for achieving the desired texture in firm yogurt; (b) stirring during fermentation may inhibit lactic acid production, potentially prolonging the fermentation process and compromising the final product's quality; and (c) stirring could introduce air into the mixture, disrupting the anaerobic conditions necessary for proper fermentation.[80–82] None of these concerns pertain to altering the flavor or aroma.

Aguirre-Ezkauriatza et al. [82] investigated the impact of mixing during yogurt fermentation. The average viscosity values from three separate stirred experiments were 158.0, 150.0, and 152 cP. In comparison, samples from the non-stirred fermentation experiments showed significantly higher viscosity values of 4088, 4052, and 4096 cP. Stirring during fermentation disrupts the formation of a uniform protein network, leading to weaker gel structures and lower viscosity in stirred samples. In contrast, non-stirred fermentation allows the protein matrix to form undisturbed, resulting in a firmer gel and much higher viscosity.

The study by Lubbers et al. [83] highlights how changes in yogurt's rheology and apparent viscosity affect flavor and aroma release. Specifically, they found that as the apparent viscosity of the yogurt increased over time (due to factors such as acidification and exopolysaccharide production), the release of aroma compounds in the headspace decreased. This reduction in flavor release was particularly noticeable for esters like methyl 2-methylbutanoate, ethyl hexanoate, and hexyl acetate. The increase in yogurt's viscosity, which results from the strengthening of the protein network and the production of exopolysaccharides, likely restricts the mobility and release of volatile aroma compounds. Consequently, the changes in texture and viscosity alter the perceived flavor and aroma of the yogurt, potentially making the product less aromatic and flavorful. This interaction between rheology and flavor release suggests that yogurt's physical properties and composition are crucial to its sensory characteristics.[84] However, in their sensory experiment[82], a 22-member non-expert panel conducted 66 yogurt triangle tests comparing samples from stirred and non-stirred fermentations. Despite the differences in viscosity, the panel did not distinguish between the samples.[82]

The cheese quality is fundamentally linked to the microorganisms used as starters for both acidification and flavor development. These microorganisms play a crucial role throughout the cheese-making process, from manufacture to ripening. During cheese production and aging, microbial activity significantly modifies all major milk components: Carbohydrates, Proteins, and Lipids. These changes are essential for developing the cheese's distinctive flavor profile. Flavor development in cheese is a dynamic biochemical process influenced by several key factors: Milk composition, Curd processing techniques, Ripening conditions, Enzymes naturally present in cow's milk, Indigenous microorganisms, or added starter cultures. Cheese microorganisms serve as the primary source of enzymes that drive flavor development. The complex flavor profile of cheese comprises various compounds: Sapid compounds, including Organic acids, Peptides, Amino acids, Added sodium chloride (NaCl), and Volatile aroma compounds. These components work in concert to create the unique sensory experience of each cheese variety. By carefully controlling these factors, particularly the selection and management of microorganisms, cheesemakers can significantly influence their final product's quality and flavor characteristics.[85,86]

One study [87] highlights that the agitation of cheese milk, mainly through churning, can lead to undesirable flavors due to the activation of milk lipase. The agitation process used in commercial cheese production often differs from experimental conditions, which may impact the results. However, vigorous agitation can cause rancid and off-flavors in cheese. To mitigate these issues, it is recommended to minimize unnecessary agitation of cheese milk at all farm stages, in transit, and at the factory before the milk is set. Lower agitation temperatures are also beneficial as they reduce the detrimental effects on cheese quality. Cooling cheese milk is advised, countering the practice of delivering milk uncooled to activate lipase. For milk transported longer distances, more cooling is necessary. [87]

Although not studying the stirring, as part of their research, Cunha et al. [88] evaluated the effect of three different types of fat (butter oil, partially hydrogenated soybean fat, and soybean oil) on the rheological properties and the sensory acceptance of spreadable cheese analogs. Their spreadable cheese analogs showed greater hardness than that manufactured with butter oil and higher values for the elastic ($G'$) and viscous ($G''$) moduli. However, the cheese manufactured with butter oil was much better evaluated in the sensory evaluation.

Cao et al. [89] determined how specific variations in process parameters could influence the metabolism of bacteria and, therefore, whether the choice of process parameters might constitute a lever to modulate the volatility of semi-hard cheeses. One of their studied processing parameters was stirring (after cutting coagulum) time on aroma formation. They reported that a reduction in mixing time selectively reduced the buttery flavor-associated compounds produced by *L. lactis* [90] and increased the amounts of some compounds associated with "aged-cheese" and "Swiss-cheese-related flavor" notes and resulted from the activities of *P. freudenreichii.* The rheological properties of the low-fat desserts were significantly affected by starch and κ-carrageenan content, leading to significant differences in the orally perceived thickness. Despite these differences, neither flavor release nor the perception of strawberry flavor intensity of the low-fat dairy dessert was significantly affected.[91]

## Conclusion

This review explores the intricate relationship between mixing dynamics, rheological properties, and the development of flavors and aromas in fermented foods, a critical aspect of food production. Mixing is a crucial physical factor in fermentation, influencing the distribution of substrates, microorganisms, and metabolites. Variations in shear forces and flow patterns affect reaction rates, mass transfer, and fermentation kinetics. Efficient mixing is essential for optimizing the release of volatile compounds and integrating flavors, significantly impacting the sensory profile of fermented products.

The rheological behavior of fermentation broths, characterized by non-Newtonian fluids and high viscosity, decisively influences biosynthesis processes. Viscosity affects mass and heat transfer, mixing intensity, cell growth rate, and product accumulation rate. The modification of rheological properties during fermentation necessitates careful management to avoid stagnant regions and ensure optimal process control. Rheological properties are closely linked to sensory attributes, including mouthfeel and flavor intensity, with increased viscosity reducing the perceived intensity of volatile and non-volatile compounds due to inefficient mixing and reduced diffusion rates.

The impact of mixing on specific fermented foods such as bread, wine, vinegar, beer, and yogurt is significant. For instance, kneading develops the gluten network in bread making and incorporates air bubbles, enhancing flavor perception through a light and airy crumb structure. In wine and beer production, mixing influences the extraction of flavor and aroma compounds from raw materials. In yogurt and cheese production, it affects the formation of key flavor compounds like diacetyl and lactones. Understanding and controlling these physical interactions is essential for optimizing fermentation to meet consumer preferences and quality standards. Applying advanced mixing techniques and a thorough understanding of rheology can lead to better flavor retention, enhanced aroma profiles, and improved product quality in the food and beverage industry. Manufacturers can better meet consumer preferences and quality standards by tailoring the sensory attributes of fermented foods through controlled mixing and rheological management. This approach enables the creation of novel, artisanal food products that maintain high nutritional and sensory qualities. Further research is needed to fully elucidate the mechanisms underlying the relationship between rheological properties and aroma release and the effects of different mixing regimes on microbial activity and substrate availability.